\documentclass[10pt,aps,prb,twocolumn,amsmath,amssymb,
longbibliography,
]{revtex4-2}
\usepackage{graphicx}
\usepackage{epsfig}
\usepackage{dcolumn}
\usepackage{bm}
\usepackage{bbm}
\usepackage[normalem]{ulem}
\usepackage{color}
\usepackage{float}
\usepackage[colorlinks=true,linkcolor=blue,citecolor=blue,urlcolor=blue,breaklinks=true]{hyperref}
\usepackage{verbatim}
\usepackage{subfigure}
\usepackage[ruled,vlined]{algorithm2e}
\usepackage{tikz}
\usepackage{braket}

\def\be{\begin{equation}}       \def\ee{\end{equation}}
\def\bea{\begin{eqnarray}}      \def\eea{\end{eqnarray}}
\def\ba{\begin{array}}
\def\ea{\end{array}}
\def\bnum{\begin{enumerate} }
\def\enum{\end{enumerate}}

\def\=>{\Rightarrow}
\def\>{\rightarrow}

\def\eye2{Fathbb{I}}

\usepackage{ listings }

\def\Eq#1{Eq.~(\ref{#1})}
\def\Fig#1{Fig.~\ref{#1}}

\newcommand{\w}{{\omega}}

\renewcommand{\>}{\rangle}

\newcommand{\eq}[2]{
	\begin{equation}
	#1 \label{#2}
	\end{equation}
}

\newcommand{\vect}[1]{\boldsymbol{#1}}

\usepackage{listings}
\usepackage{color}
\definecolor{lightgray}{gray}{1}

\newtheorem{theorem}{Theorem}
\newtheorem{corollary}{Corollary}[theorem]

\newcommand\COMMENTED[1] {}

\begin{document}

\title{Addressing the Infinite Variance Problem in Fermionic Monte Carlo Simulations: \\ Retrospective Error Remediation and the Exact Bridge Link Method}
\author{Zhou-Quan Wan}
\email{zwan@flatironinstitute.org}
\affiliation{Center for Computational Quantum Physics, Flatiron Institute, New York, NY 10010, USA}

\author{Shiwei Zhang}
\email{szhang@flatironinstitute.org}
\affiliation{Center for Computational Quantum Physics, Flatiron Institute, New York, NY 10010, USA}

\begin{abstract}
We revisit the infinite variance problem in fermionic Monte Carlo simulations, which is widely encountered in areas ranging from condensed matter to nuclear and high-energy physics.
The different algorithms, which we broadly refer to as determinantal quantum Monte Carlo (DQMC), are applied in many situations and differ in details, but they share a foundation in field theory, and often involve fermion determinants whose symmetry properties make the algorithm sign-problem-free.
We show that the infinite variance problem arises as the observables computed in DQMC tend to form heavy-tailed distributions.  
To remedy this issue retrospectively, we introduce a tail-aware error estimation method to correct the otherwise unreliable estimates of confidence intervals.
Furthermore, we demonstrate how to perform DQMC calculations 
that eliminate the infinite variance problem for a broad class of observables.
Our approach is an exact bridge link method, which involves a simple and efficient modification to the standard DQMC algorithm. 
The method introduces no systematic bias and is straightforward to implement with minimal computational overhead. 
Our results establish a practical and robust solution to the infinite variance problem, with broad implications for improving the reliability of a variety of fundamental fermion simulations.
\end{abstract}

\date{\today}
\maketitle

\section{Introduction} 
Quantum Monte Carlo (QMC) methods are among the most powerful tools for studying strongly correlated quantum systems. 
Within this family, the determinantal quantum Monte Carlo (DQMC) method~\cite{BSS1981} provides a numerically exact approach for simulating interacting fermions.
For many models of physical interest in condensed matter to nuclear and high-energy physics, 
this method turns out to be free of the sign problem
\cite{congjun2005prb,wanglei2015prl,Li2016PRL, wei_majorana_2016,wei2018semigroup, ZXLiQMCreview}.
Numerous applications exist in an extensive literature spanning several decades, a small fraction of which include studies of critical behavior near phase transitions \cite{Otsuka2016prx,Erez2017pnas,li2017fiqcp,zixiang2018,xxy2021prl,wan2022prb,ziyang2023gny,erramilli2023gross}, zero-density lattice QCD \cite{rmp2010}, and direct comparisons with cold-atom or condensed matter experiments \cite{shihao2015pra,shihao2016prl,Paiva2010prl,Peter2017science,mingpu2022annualreview,yuanyao2025prl,chunhan2025arxiv}.
In these cases, DQMC serves as a valuable non-perturbative tool for extracting physical insights and providing accurate, reliable benchmarks for both experiment and theory.
Despite its success and wide applicability, DQMC suffers from a subtle yet critical statistical pathology,  
the infinite variance problem~\cite{shihao2016}. 
The issue arises because certain observables diverge in the vicinity of zeros in the sampling probability distribution.
As a result, conventional error estimation techniques become unreliable.
This can be particularly catastrophic when the sign problem is absent and the result is supposedly numerically exact.

The infinite variance problem is not unique to DQMC; it also arises in various other Monte Carlo methods. 
For example, although standard observables in variational Monte Carlo (VMC) typically avoid infinite variance, they can exhibit anomalous distributions that impede statistical convergence~\cite{Trail2008}. 
Furthermore, quantities such as gradients or the Fisher information matrix — used in stochastic reconfiguration (SR) methods \cite{Sorella1998prl_sr,Stephan2023scipost} — can suffer from the infinite variance problem.
Similar problems also emerge in the calculation of forces within both diffusion Monte Carlo (DMC) and VMC frameworks \cite{ASmethod,Wagner2020regularization}, as well as in time-evolution simulations performed using the variational principle \cite{Filippo2023tVMC}. 
The infinite variance problem has also attracted attention in the machine learning community, particularly in the context of stochastic gradient descent (SGD)-based optimization methods \cite{simsekli2019tail,lingjiong2021heavytailsgd}.
These problems share a common feature with the infinite variance 
problem in DQMC, namely the presence of nodes in the sampling distribution.

As the demand for high-precision simulations continues to grow, accurate interpretation of Monte Carlo uncertainties has become increasingly critical. 
This is particularly true in studies of quantum critical behavior, where numerical results are compared with theoretical predictions such as perturbation methods or the conformal bootstrap \cite{Otsuka2016prx,Erez2017pnas,li2017fiqcp,zixiang2018,xxy2021prl,wan2022prb,ziyang2023gny,erramilli2023gross}. 
Equally important are comparisons with experimental measurements—for example, benchmarking lattice QCD calculations against high-energy experimental data \cite{rmp2010}, extracting the equation of state in strongly correlated systems, investigating the BCS-BEC crossover in the two-dimensional Fermi gas \cite{shihao2015pra,shihao2016prl}, or comparing with experimental results from ultracold atom quantum simulators of the Hubbard model for benchmarking and thermometry \cite{Paiva2010prl,Peter2017science,mingpu2022annualreview,yuanyao2025prl,chunhan2025arxiv}.
In all these contexts, the reliability of Monte Carlo error estimates directly impacts the ability to draw meaningful physical conclusions.
Addressing the infinite variance problem is thus essential for advancing the role of numerical simulation in both theoretical and experimental frontiers of modern physics.

In Ref.~\cite{shihao2016}, the bridge link method was proposed as an effective solution to the infinite variance problem. 
By inserting an additional time slice into the imaginary-time path integral, the auxiliary fields are sampled from a modified distribution that is free of zeros, thereby eliminating the divergence in observable estimators. 
However, the original and subsequent~\cite{Alexandru2023prd} implementations required a secondary Monte Carlo procedure and do not provide a perfect solution. 
Alternative approaches~\cite{assaad2022spikes,Yunus2022prd} also each have significant limitations. 

In this work, we present a detailed analysis of the infinite variance problem, giving an understanding of statistical errors in its presence. 
For previously obtained data affected by infinite variance, we propose a simple and effective method to retrospectively correct error estimates using information about the tail of the distribution. 
More importantly, we propose a practical solution that overcomes the limitations of previous approaches. 
Our method, which we refer to as the exact bridge link method, builds upon the core idea of the original bridge link approach. 
By leveraging the sign-problem-free properties of the model, we demonstrate that the infinite variance problem can be completely eliminated for a broad class of observables. 
The proposed method is general, easy to implement, and incurs minimal computational overhead. Crucially, it does not rely on approximations or introduce any systematic errors.

The remainder of this paper is organized as follows. 
Section~\ref{sec:retrospective_remediation} presents the retrospective correction method for better estimates of statistical confidence levels. 
Representative QMC simulation results are shown to illustrate the manifestations and consequences of the infinite variance problem.
Then we introduce the method for correcting confidence intervals a posteriori from existing DQMC simulations, assuming knowledge of the tail behavior of the distribution.
Section~\ref{sec:eliminating_ivp} presents our exact bridge link method. After reviewing the DQMC algorithm and analyzing the underlying origin of the infinite variance problem within this formalism, we describe the exact bridge link method and provide a rigorous demonstration of its effectiveness in eliminating infinite variance across a broad class of observables.
We further evaluate its scalability, benchmark its performance, and apply it to the calculation of charge-$4e$ correlations to showcase its power in accessing previously inaccessible observables.
Section~\ref{sec:discussion} concludes with a discussion of the broader implications of our findings and outlines open questions and future directions.

\section{Retrospective remediation of the infinite variance problem}
\label{sec:retrospective_remediation}

In this section, we demonstrate the infinite variance problem by examining representative QMC data where this issue arises.
We do not delve into the details of the QMC algorithm or the underlying causes of the infinite variance here—these aspects will be addressed in Section~\ref{subsec:dqmc_and_ivp}.
Instead, our focus here is on the practical consequences: 
how the underlying distribution that gives rise to infinite variance shows a heavy-tailed behavior, 
and how this impacts statistical error estimation in simulations. 
To address this, we introduce a tail-aware error estimation method that enables retrospective correction of confidence intervals, assuming prior knowledge of the distribution’s tail behavior.

\subsection {Infinite variance problem and heavy-tailed distribution}
\label{subsec:infinite variance}

\begin{figure}
    \centering
    \includegraphics[width=1.0\linewidth]{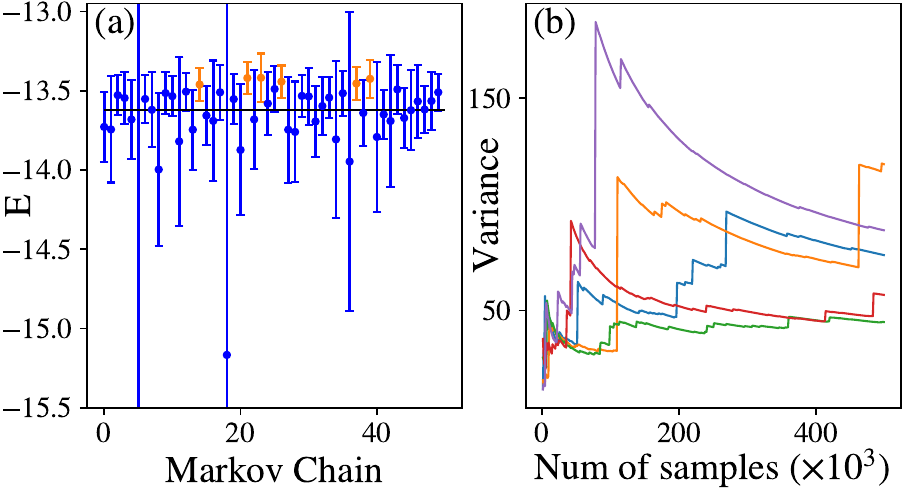}
    \caption{(a) Energy estimates from 50 independent DQMC runs. Error bars indicate twice the SEM, corresponding to a conventional $\sim $95.4\% confidence interval.     
    Orange points denote cases where the true energy value (black horizontal line, obtained using the exact bridge link method described in Sec.~\ref{sec:eliminating_ivp}) lies outside the estimated confidence interval. 
    (b) Variance of the energy estimator versus sample size for the first 5 independent runs, with each color indicating one run.}
    \label{fig:energy_without_bridge}
\end{figure}

We begin by examining typical QMC simulation data. 
In \Fig{fig:energy_without_bridge}, we present results from 50 independent DQMC simulations of the half-filled Hubbard model with interaction strength $U/t=4$ on a $4 \times 4$ periodic lattice. Specifically, we used $\Delta\tau = 0.05$ and 
a projection time $\Theta = 20$ in these simulations. 
The error bars correspond to twice the standard error of the mean (SEM). 
Each estimate uses the final $5 \times 10^5$ samples from a total of $10^6$, discarding the initial half for thermalization. 
SEMs are computed via the rebinning method with a bin size of 100  to account for autocorrelation effects.

Under standard statistical assumptions, the error bars correspond to an approximately 95.4\% confidence interval for the sample mean. 
However, as \Fig{fig:energy_without_bridge}(a) shows, in 6 out of 50 independent runs, the true energy value lies outside the reported confidence interval—almost three times the expected $\sim$4.6\% failure rate. 
It is also noteworthy that several runs exhibit significantly larger error bars compared to others ---
even though they all use the same number of samples ---
and in those cases, the sample mean deviates substantially from the true value.
As illustrated in \Fig{fig:energy_without_bridge}(b), the sample variance $s^2 \equiv \frac{1}{N-1}\sum_{i=1}^N (x_i-\bar x)^2  $ fails to converge with increasing sample size $N$, but instead grows erratically, characteristic of the {\it{infinite variance}} problem.

\begin{figure}
    \centering
    \includegraphics[width=1.0\linewidth]{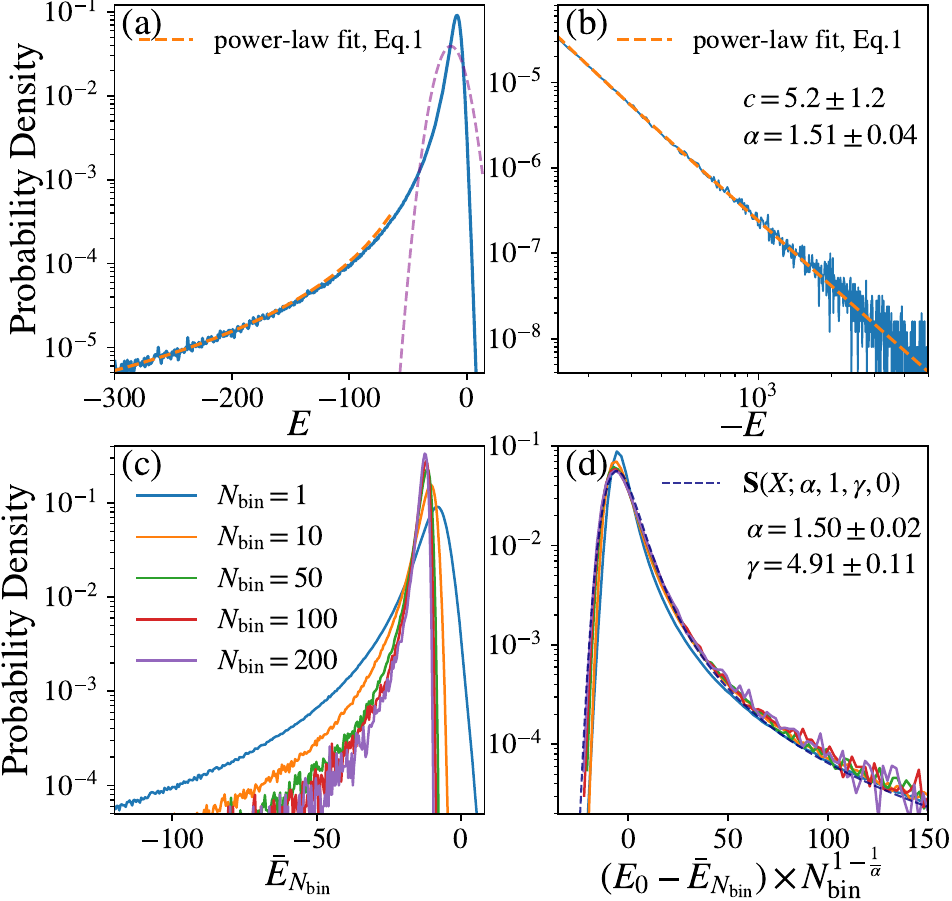}
    \caption{Characteristics of the probability distribution from DQMC data.
    (a) Probability density of raw energy measurements taken at the middle time slice of each sweep,
    from the dataset in \Fig{fig:energy_without_bridge}(a). The purple curve shows a normal distribution $\mathcal{N}(\bar{E}, \sigma )$ for comparison, with $\sigma = \sqrt\frac{\pi}{2}\cdot\overline{|E-\bar E|}$, chosen to match the mean absolute deviation of the data since the variance diverges in this case.
    The dashed orange line is the fitted power-law model from panel (b). 
    (b) Tail distribution of the energy data, which exhibits a power-law decay. 
    The dashed orange line displays the
    result from the fit using \Eq{eq:power-fit-tail}.
    (c) The probability density of bin-average data of energy 
    for different bin size $N_{\text{bin}}$.
    The spread of $\bar E_{N_\text{bin}}$ appears to decrease as $N_{\text{bin}}$ is increased, but not as rapidly as $1/\sqrt{N_{\text{bin}}}$, and the distribution does not approach a Gaussian.
    (d) Data collapse of all $\bar E_{N_\text{bin}}$ data
    onto a stable distribution $\mathbf{S}(X;\alpha, 1, \gamma, 0)$. The result of the fit using this form, shown by the blue dashed line, is consistent with that from the fit in panel (b).
    }
    \label{fig:distribution}
\end{figure}

The problem becomes even more evident if we examine the full distribution of the DQMC data.
As shown in \Fig{fig:distribution}(a), 
the distribution exhibits a pronounced long tail extending toward negative infinity, in stark contrast to the rapid decay of the normal distribution.
To quantitatively assess the tail behavior, we present the distribution on a log-log scale in \Fig{fig:distribution}(b), where a clear linear trend emerges, indicative of power-law scaling.
Fitting the tail with the power-law form 
\eq{
y = \frac{\alpha c}{(x-x_0)^{\alpha + 1}},
}{eq:power-fit-tail}
we extract a characteristic exponent 
$\alpha = 1.51\pm 0.04$ and a tail coefficient $c = 5.2\pm 1.2$, which is displayed in \Fig{fig:distribution}(b).
A distribution with a power-law tail is commonly referred to as heavy-tailed, with the parameter $\alpha$ known as the characteristic exponent.
When $1<\alpha<2$, the distribution has a finite mean but divergent variance, while for $0<\alpha\leq 1$ even the mean is undefined.

The present data falls into the $1<\alpha<2$ regime.
A direct consequence of this is the breakdown of the central limit theorem (CLT), which requires finite variance as a prerequisite.
As illustrated in \Fig{fig:distribution}(c), the distribution of the sample mean, defined as 
$\bar X_{N_\text{bin}}\equiv \frac 1 {N_\text{bin}}\sum_{i=1}^{N_\text{bin}} X_i$,
does not converge to a normal distribution, even for large ${N_\text{bin}}$.
This breakdown of the CLT is precisely why the conventional standard error estimator fails:
without finite variance, the usual relationship between sample size and uncertainty no longer holds, leading to unreliable or misleading confidence intervals. 

Infinite variance does not necessarily imply that the sample mean ceases to be a valid estimator of the true mean. 
As long as the underlying distribution possesses a finite mean — as is the case for $\alpha > 1$ — the law of large numbers (LLN) remains applicable. 
As shown in Fig.~\ref{fig:distribution}(c), the fluctuations of the sample mean do appear to diminish with increasing sample size.
In the context of heavy-tailed distributions with power-law behavior, the generalized central limit theorem (GCLT) governs the asymptotic behavior of the sample mean. 
Here we briefly summarize its key implications; further details can be found in the Appendix and Refs.~\cite{gnedenko1968limit,nolan2020univariate}.
If a distribution $\mathbb P(X)$
satisfies the following asymptotic tail conditions:
\eq{
\begin{array}{ll}
\lim_{x\rightarrow\infty} x^{\alpha}\mathbb P(X>x)&=c^+, \\
\lim_{x\rightarrow\infty} x^{\alpha}\mathbb P(X<-x)&= c^-,
\end{array}
}{}
where $\mathbb P(X>x)$ and $\mathbb P(X<-x)$ denote the cumulative probabilities that $X$ exceeds $x$ and is less than $-x$, respectively. 
Assume further that $1<\alpha<2, c^-\geq 0, c^+\geq 0$ and $c^-+c^+>0$.
Then, the GCLT states that the properly rescaled sample mean converges in distribution to a stable law \cite{lévy1925calcul,gnedenko1968limit,nolan2020univariate}:
\eq{
N^{1-\frac 1 \alpha}\bar X_N \stackrel{d}{\rightarrow} \mathbf{S}(X;\alpha,\beta,\gamma, \delta),
}{GCLT}
where $\mathbf{S}(X; \alpha,\beta,\gamma, \delta)$ 
denotes a stable distribution of $X$ (see Appendix for its definition). 
Except for the location parameter $\delta=\mathbb E[ X]$,  all other parameters of the limiting distribution are fully determined by the tail behavior of the original distribution: the characteristic exponent $\alpha$; the skewness parameter  $\beta=\frac{c^+-c^-}{c^++c^-}$, reflecting the asymmetry of the tail coefficients $c^\pm$; the scale parameter $\gamma = \left(\frac{2\Gamma(\alpha) \sin(\pi \alpha/2)}{\pi (c^++c^-)}\right)^{-1/\alpha}$.

As shown in \Fig{fig:distribution}(d), the distributions of sample means at various sample sizes collapse onto a universal curve when rescaled by the factor $N^{1-\frac 1 \alpha}$, in agreement with the predictions of the GCLT.
Fitting the rescaled data to a stable distribution yields a characteristic exponent $\alpha = 1.50 \pm 0.02$ and a scale parameter $\gamma = 4.91 \pm 0.11$.
(In the fit, the skewness parameter is fixed at $\beta = 1$, reflecting the one-sided nature of the distribution.)
These results are consistent with the GCLT prediction 
of $\alpha$ and $c$ (i.e., $c^++c^-$) based on the tail coefficients shown in \Fig{fig:distribution}(b).  

\subsection{Tail-aware error estimation}
\label{subsec:error_estimation}

Based on the scaling behavior of the sample mean described in \Eq{GCLT}, the statistical uncertainty (or ``error") of the sample mean must scale as $N^{\frac{1}{\alpha} - 1}$, in contrast to the conventional $1/\sqrt{N}$ scaling that holds when the variance is finite.
We now turn to the challenge of estimating confidence intervals for data drawn from distributions with infinite variance.
Recall that the conventional approach estimates the SEM as
\eq{
\text{SE}_N(X) \equiv \sqrt{\frac{s^2}{N}} = \sqrt{\frac {\sum_{i=1}^N(X_i-\bar X_N)^2} {N(N-1)}  }.
}{}
If the CLT holds, this estimator converges to $\sigma / \sqrt{N}$ in the limit $N \rightarrow \infty$, where $\sigma$ is the variance of the underlying distribution. 
Multiplying the SEM by two yields an approximate $95.4\%$ confidence interval for the sample mean — commonly referred to as the two-sigma confidence interval.
However, for heavy-tailed distributions with $1<\alpha<2$, these conventional expressions become invalid because the variance $\sigma^2$ is formally divergent.
Counterintuitively, though, $\text{SE}_N$ does not diverge as the sample size goes to infinity.
In the heavy-tailed case, one can show that the appropriately rescaled SEM converges in distribution to a nontrivial limiting form:
\eq{{N^{1-\frac{1}{\alpha}}\text{SE}_N(X)} \stackrel{d}{\rightarrow} \sqrt{\eta_{\frac \alpha 2}(c^-+c^+)Z'}, }{standard_error}
where $Z'\sim\mathbf{S}(\alpha/2,1,1,0)$, $\eta_\alpha(c) = \left(\frac{2\Gamma(\alpha)\sin(\frac{\pi\alpha}{2})}{\pi c}\right)^{-1/\alpha}$.
Comparing \Eq{GCLT} and \Eq{standard_error}, we observe that $\text{SE}_N$ exhibits the same scaling behavior as the sample mean itself—namely, $N^{\frac 1 \alpha-1}$. 
This implies that, although the variance diverges, the finite-sample SEM remains finite and approximately captures the correct scaling of statistical fluctuations.
It does not retain the same interpretation as in the finite-variance case, however, and no longer corresponds to the conventional confidence level.

A better estimate of the confidence level is possible 
for heavy-tailed distributions. 
One example is the $m$-out-of-$n$ bootstrap method, which constructs asymptotically correct confidence intervals, provided that the resample size satisfies $m/n\rightarrow0$ as $m, n \to \infty$ \cite{Arcones1989,hall1996bootstrap}.
We have tested this method on a number of systems with data generated from DQMC calculations, and 
recommend its use in practical scenarios involving heavy-tailed data — particularly when the tail parameters are unknown or difficult to estimate reliably.
(It is worth emphasizing that, without sufficiently large sample size, this method can lead to unreliable confidence interval estimates, so application requires extra caution.)
With this work we provide scripts 
for a simple implementation of $m$-out-of-$n$ bootstrap method~\footnote{See \href{https://github.com/Zhouquan-Wan/error-estimation-with-infinite-variance}{https://github.com/Zhouquan-Wan/error-estimatio\-n-with-infinite-variance}\,.}. 

\begin{figure}
    \centering
    \includegraphics[width=1.0\linewidth]{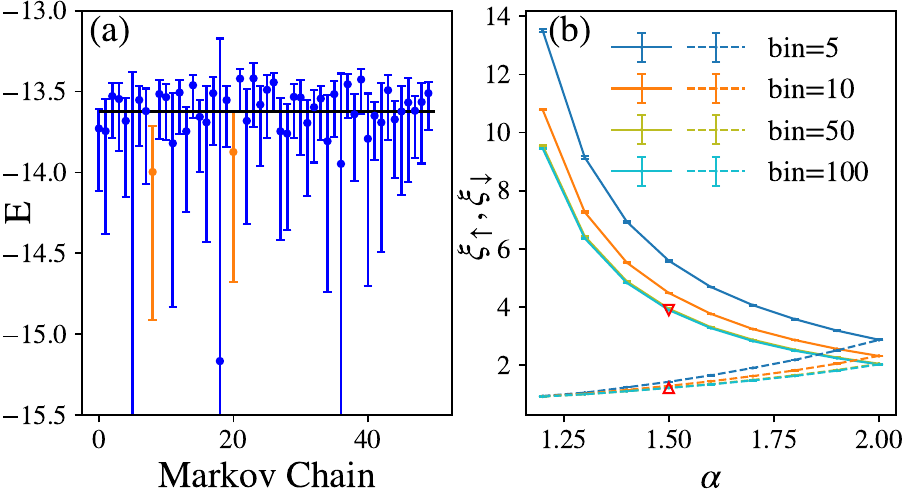}
    \caption{Error remediation incorporating tail information.
    (a) Tail-aware error estimation for the same dataset shown in \Fig{fig:energy_without_bridge}(a). 
    The confidence interval is constructed by rescaling the original SEM estimate, $\text{SE}_N$.
    The interval is given by $(\bar{E} - \xi_\downarrow\, \text{SE}_N, \bar{E} + \xi_\uparrow\, \text{SE}_N)$, with $\xi_\uparrow = 1.22$ and $\xi_\downarrow = 3.90$ which are obtained following the method outlined in the text.
    Out of the 50 data points, 2 (indicated in orange) are outside the corrected interval, consistent with the target confidence level.
    (b) The scaling factors $\xi_\uparrow$ (dashed lines) and $\xi_\downarrow$ (solid lines) as functions of the characteristic exponent $\alpha$, corresponding to the $2\sigma$ confidence level. 
    The symbols show the final $\xi_\uparrow$ and $\xi_\downarrow$ values used in panel (a). 
    }
    \label{fig:error_estimation}
\end{figure}

A simple strategy can be developed to remediate the error estimation from the SEM, even for small bin sizes, 
if the tail parameters $\alpha,\beta$ are known.
Suppose we have $N\times M$ data points drawn from a heavy-tailed distribution, partitioned into $N$ bins with size $M$. 
By computing the average within each bin, we obtain $N$ approximately independent samples drawn from the corresponding stable distribution as long as $M$ is sufficiently large.
We can then calculate the SEM from these $N$ samples. 
To construct a meaningful confidence interval, we introduce two scaling factors, $\xi_{\downarrow}$ and $\xi_{\uparrow}$, such that the interval $(\bar X-\xi_{\downarrow}\text{SE}_N,\bar X+\xi_{\uparrow}\text{SE}_N)$ achieves the desired confidence level — for example, $95.4\%$.
These scaling factors depend only on the limiting stable distribution, which is fully characterized by the characteristic exponent $\alpha$ and the skewness parameter $\beta$.
Given knowledge of $\alpha$ and $\beta$, we can numerically generate samples from the corresponding stable distribution and determine $\xi_{\downarrow}$ and $\xi_{\uparrow}$ using Monte Carlo methods. 
The procedure is as follows (our actual implementation is provided at \cite{Note1}):
\begin{enumerate}
    \item Generate $N_g \times N$ independent samples from the stable distribution $\mathbf{S}(\alpha, \beta, 1, 0)$, and partition them into $N_g$ groups of size $N$.

    \item For each group $i = 1, \dots, N_g$, compute the studentized sample mean $T_N(i) = \text{E}_N(i)/\mathrm{SE}_N(i)$ to construct the empirical distribution.
    
    \item From the sorted $T_N(i)$, extract the lower and upper quantiles at levels $\epsilon/2$ and $1 - \epsilon/2$, respectively. These quantiles estimate the scaling factors $\xi_{\downarrow},\xi_{\uparrow}$.
\end{enumerate}
By repeating the process, we can give an estimation of the scaling factors, which can then be combined with the original $\text{SE}_N$ results to obtain the remediated confidence interval. 

In \Fig{fig:error_estimation}(a), we present the results of the remediated SEM-based error estimation applied to the same dataset from \Fig{fig:energy_without_bridge}.
The rescaling factors are obtained by Monte Carlo sampling using \texttt{scipy}~\cite{2020SciPy-NMeth}, from a one-sided $\alpha$-stable distribution with characteristic exponent $\alpha = 1.5$.
Unlike the conventional 2-SEM error bars shown in \Fig{fig:energy_without_bridge}(a), the rescaled error bars exhibit significantly improved consistency with the target confidence level. 
Moreover, the asymmetry of the resulting confidence intervals accurately reflects the skewness of the underlying distribution.

In \Fig{fig:error_estimation}(b), we illustrate the dependence of the rescaling factors $\xi_{\uparrow}$ and $\xi_{\downarrow}$ on the characteristic exponent $\alpha$ and the number of bins $N$.
As evident from the plot, both factors converge to asymptotic values that depend solely on $\alpha$ (with $\beta$ fixed) in the limit $N \to \infty$.
As $\alpha$ approaches 2, these rescaling factors recover the conventional value of 2, consistent with the standard case of finite variance under the CLT.
For $\alpha<2$, however, the rescaled error bars exceed the conventional ones, reflecting the underestimation of statistical uncertainty when using standard error estimation in the presence of heavy-tailed distributions.
Even then, when $1<\alpha<2$, the distribution of the studentized sample mean converges to a limiting distribution with Gaussian-like tails~\cite{Logan1973tlimit}.
This helps to explain why results in the literature 
from calculations such as DQMC,  with infinite variance, can seem ``reasonable'' even with error estimates based on the ``standard'' approach.
The true confidence intervals differ primarily by a scaling factor, which itself converges as the confidence level increases.

This rescaling approach offers a practical framework for re-evaluating existing QMC data affected by the infinite variance problem.
Once the tail behavior of the distribution is characterized, the confidence interval can be reliably estimated and revised by applying appropriate multiplicative factors to the SEM.
In Section~\ref{subsec:dqmc_and_ivp}, we provide insights into the typical values of the characteristic exponent $\alpha$ in DQMC, based on the analysis presented in Ref.~\cite{assaad2022spikes}. 
The behavior of $\alpha$ has also been investigated in the context of hybrid Monte Carlo simulations \cite{ulybyshev2017_powerlaw}.

To conclude this section, we highlight several important implications of infinite variance.
First, in many cases, the statistical fluctuations of the sample mean scale as $N^{\frac{1}{\alpha}-1}$, which can be significantly slower than the conventional $1/\sqrt{N}$ scaling in finite-variance cases and can change the complexity of QMC algorithms.
Given that typical simulations may involve up to $10^6$ samples or more, this slow convergence can pose a serious limitation in practice.
Second, conventional error estimation methods—such as the SEM—can systematically underestimate the true uncertainty in the presence of infinite variance, especially when the underlying distribution is skewed, resulting in misleading conclusions.
Moreover, the sample mean itself converges to a heavy-tailed distribution, implying that commonly used techniques for error propagation or uncertainty estimation (e.g., least-squares fitting or naive bootstrap methods) likely also yield inaccurate results when applied to QMC data with heavy-tailed characteristics.
These issues underscore the critical need to address the infinite variance problem when performing practical Monte Carlo simulations.
As we discuss next, with minimal adjustment in the algorithm/code, it turns out that the infinite variance problem can be completely removed.

\section{Eliminating the infinite variance problem via the exact bridge link method}
\label{sec:eliminating_ivp}
In this section, we begin by introducing the background of the DQMC algorithm and re-examining the origin of the infinite variance problem within this framework.
We then present the \textit{exact bridge link method}, a simple yet general modification that is easy to implement and effectively eliminates the infinite variance problem for a broad class of observables. 
To demonstrate its practical utility, we apply the method to compute charge-$4e$ correlations in the attractive SU(4) Hubbard model.

\subsection{The DQMC algorithm and the origin of the infinite variance problem}\label{subsec:dqmc_and_ivp}

For concreteness, we focus on the zero-temperature version of the DQMC algorithm; however, the discussion regarding infinite variance also applies to the finite-temperature formulation.
In the zero-temperature DQMC approach, the ground-state wavefunction of the Hamiltonian $\hat H$ is obtained through imaginary-time projection:
\eq{
\ket{\psi_G}=\lim_{\Theta\rightarrow \infty} e^{-\frac \Theta 2 \hat H}\ket{\psi_T},
}{}
where $\ket{\psi_T}$ is a chosen trial wavefunction, typically taken to be a Slater determinant.
The expectation value of the observable $\hat O$ is calculated as 
\eq{\braket{\hat O} = \frac{\braket{\psi_T|e^{-\frac \Theta 2\hat H}\hat Oe^{-\frac \Theta 2\hat H}|\psi_T}}{\braket{\psi_T|e^{-\Theta\hat H}|\psi_T}}.}{}
However, the imaginary-time evolution operator $e^{-\Theta\hat H}$ is a many-body operator and cannot be directly applied to the trial wavefunction in an efficient or tractable manner. 
To address this, DQMC employs a Trotter decomposition to discretize imaginary-time evolution into $L_\tau = \Theta / \Delta \tau$ time slices. Then, a Hubbard–Stratonovich (HS) transformation is used to decouple the two-body interaction terms. This leads to the following decomposition:
\eq{
e^{-\Theta \hat H} = \prod_{i=1}^{L_\tau} e^{-\frac {\Delta\tau} 2 \hat K} \sum_{\vect x_i} e^{\hat v(\vect x_i)}e^{-\frac {\Delta\tau} 2  \hat K} + O(\Delta \tau^2),
}{}
where $\hat H = \hat{K}+\hat V$ with $\hat K$ being the one-body part (e.g.~kinetic term, $\hat V$ is the two-body or interaction term, and $\hat{v}(\vect x_i)$ the one-body operator resulting from the HS transformation at time slice $i$ with auxiliary field configuration $\vect{x}_i$.
The summation over the HS fields satisfies
$\sum_{\vect{x}_i} e^{\hat{v}(\vect{x}_i)} = e^{-\Delta \tau \hat{V}}$.

Ignoring the Trotter error due to finite time-step size $\Delta\tau$, 
the ground-state wavefunction in DQMC can be interpreted as a superposition of Slater determinants defined over the configuration space of auxiliary fields. Specifically, we can write: 
$\ket{\psi_G} \simeq \sum_{\vect x_L} \ket{\phi(\vect x_L)}$ where each configuration $\vect x_L=\{\vect x_1,\cdots, \vect x_{L_\tau/2} \} $ 
defines a Slater determinant $\ket{\phi(\vect x_L)} =\prod_{i=1}^{L_\tau/2} e^{-\frac {\Delta\tau} 2 \hat K} e^{\hat v(\vect x_i)}e^{-\frac {\Delta\tau} 2  \hat K}\ket{\psi_0} $.
Similarly for the right side: $\ket{\psi_G} \simeq \sum_{\vect x_R} \ket{\phi(\vect x_R)}$, with $\vect x_R=\{\vect x_{L_\tau/2+1},\cdots, \vect x_{L_\tau} \}$.
Then 
\eq{
\braket{\hat O}=\frac{\sum_{\vect x_L,\vect x_R} \bra{\phi(\vect x_L)}\hat O \ket{\phi(\vect x_R)}}{\sum_{\vect x_L,\vect x_R} \braket{\phi(\vect x_L)|\phi(\vect x_R)}} = \frac{\sum_{\vect x} w(\vect x)O_{\text{loc}}(\vect x)}{\sum_{\vect x}w(\vect x)},
}{O-est}
where $w(\vect x)\equiv \braket{\phi(\vect x_L)|\phi(\vect x_R)} $ is the Boltzmann weight that can be calculated by a determinant, and 
a local observable $O_{\text{loc}}$ is defined as 
\eq{O_{\text{loc}}(\vect x)\equiv \frac{\bra{\phi(\vect x_L)}\hat O \ket{\phi(\vect x_R)}}{ \braket{\phi(\vect x_L)|\phi(\vect x_R)}}. }{eq:def-loc}
In the actual simulation, the expectation value $\braket{\hat O}$ is calculated using $\braket{O_\text{loc}}_{\vect x\sim w(\vect x)}$,
where the notation ${x\sim w(\vect x)}$ means evaluating the expectation using samples of $\vect x$ drawn from $w(\vect x)$.

It can be shown that $w(\vect x)\geq 0 $ if $\hat K $ and $\hat v$ have certain symmetries or special mathematical structures \cite{congjun2005prb,wanglei2015prl,Li2016PRL, wei_majorana_2016,wei2018semigroup, ZXLiQMCreview}.
In such cases, Monte Carlo methods can be straightforwardly applied to sample the auxiliary fields, by treating $w(\vect x)$ as a probability distribution. 
This class of QMC is often referred to as 
sign-problem-free, which we focus on exclusively in this work, as the infinite variance problem is the dominant, and often only,  
uncertainty in these calculations, which can otherwise 
serve as benchmarks and provide unambiguous numerical information.

Next, to make the ensuing discussions more concrete, we delve into 
the specifics of the Hubbard model simulation that produced the data presented in Section~\ref{subsec:infinite variance}. 
The Hamiltonian of the Hubbard model is given by
\eq{
\hat H = -t\sum_{\left<i,j\right>,\sigma} (c_{i\sigma}^\dagger c_{j\sigma} + \text{h.c.}) + U \sum_i \hat n_{i\uparrow} \hat n_{i\downarrow},
}{}
where $i$ is an integer that runs through the lattice 
and labels the sites, 
$\left<i,j\right>$ denotes a sum over nearest-neighbor sites, $t$ is the hopping amplitude (set to 1 throughout this paper), and $U$ is the onsite Hubbard interaction strength. The index $\sigma = \uparrow, \downarrow$ labels the spin.
For the interaction term $U>0$, there is an exact discrete HS transformation \cite{Hirsch1983hhs}:
\eq{
e^{-\Delta\tau U(\hat n_{\uparrow}-\frac 1 2)(\hat n_{\downarrow}-\frac 1 2)}=\frac{e^{\Delta\tau U/4}}{2} \sum_{x = \pm 1}e^{i\lambda x(\hat n_\uparrow + \hat n_{\downarrow} -1)},
}{eq:HHS}
where $\lambda = \arccos(e^{-\Delta\tau U/2})$. 
Since the auxiliary fields couple to the charge density operator, we refer to this as the charge channel.
Under a partial particle-hole transformation on the spin-up fermions, $c_{i\uparrow} \rightarrow (-1)^i c_{i\uparrow}$, the repulsive Hubbard model is mapped to the attractive one. 
Correspondingly, the charge operator is transformed into a spin operator, and we refer to this as the spin channel.
These two channels involve complex numbers and are often used to compute correlation functions. 
However, they are more affected by the infinite variance problem, especially in the energy calculation.
Thus, unless explicitly stated otherwise, we use the charge channel for repulsive interactions and the spin channel for attractive interactions in this work.
At half-filling, after applying the HS transformation, the system decouples into two separate parts for spin-$\uparrow$ and spin-$\downarrow$ fermions. 
Due to the partial particle-hole symmetry, $w_\downarrow(\vect x) = w^*_\uparrow(\vect x) $. The Boltzmann weight takes the form $w(\vect x) = w_{\uparrow}(\vect x)w_{\downarrow}(\vect x) =|w_{\uparrow}(\vect x)|^2$, which is positive semi-definite and thus sign-problem-free. 
For the HS decomposition used in \Eq{eq:HHS}, $w_\uparrow(\vect x)$ can be made real by equally distributing the phase factor $\exp{(-i\lambda)}$ between the two spin sectors.

\begin{figure}
    \centering
    \includegraphics[width=1.0\linewidth]{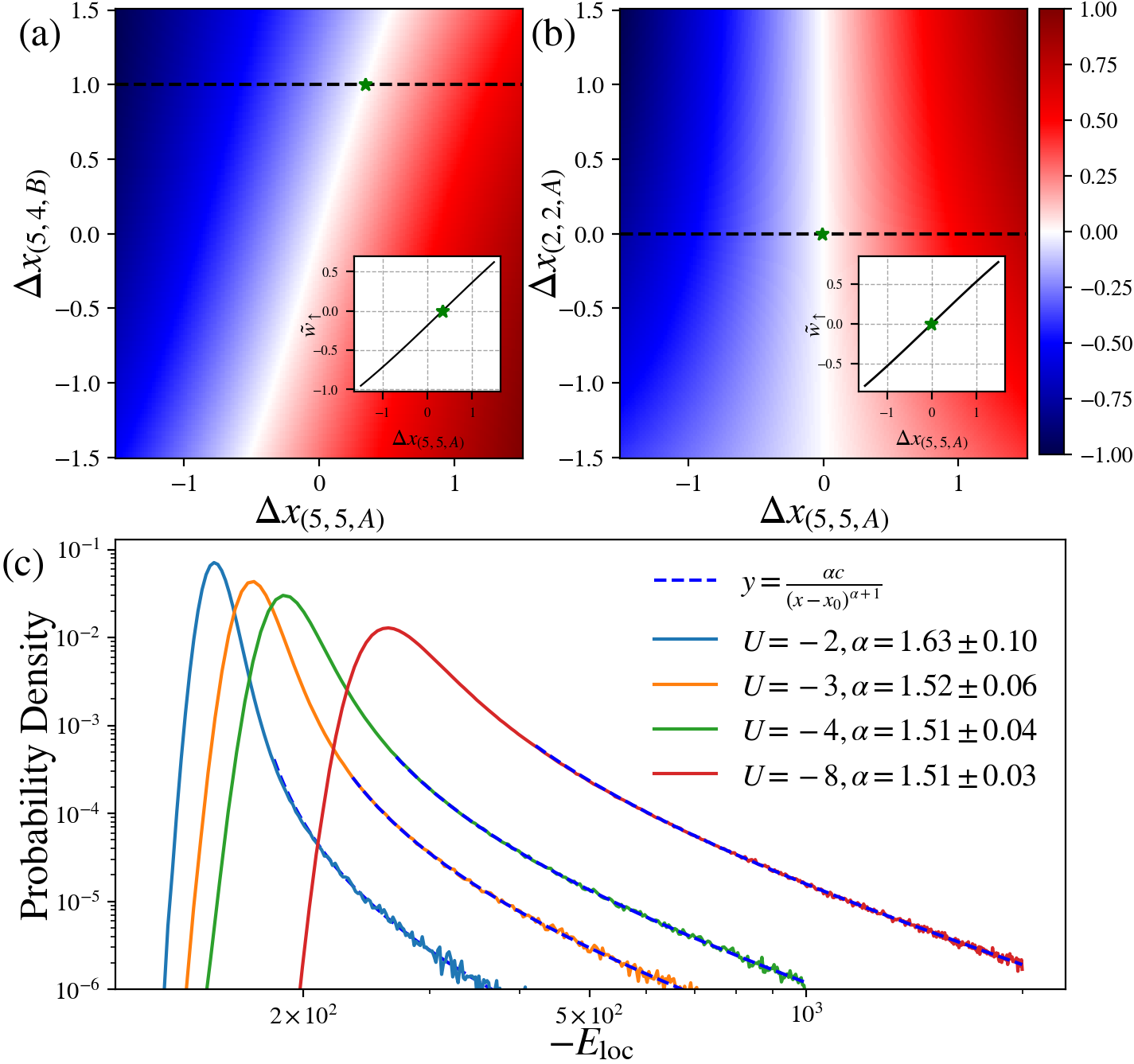}
    \caption{Nodal structure of the spin-resolved weight and local energy distribution. 
    Panels (a) and (b) show $\w_{\uparrow}(\vect x)$ as colormaps in two planar cuts in the space of auxiliary-field configurations. 
    A configuration is first identified by detecting a sign change during the simulation. 
    The last update causing the sign change occurred with the particular field $x_i$, which is shown as the horizontal axis in both panels ($\Delta x_i $ represents the displacement from the node position).
    In (a), the plane is $\Delta x_i$-$\Delta x_j$, where site $j$ is a near-neighbor of $i$. 
    In (b), the plane is $\Delta x_i$-$\Delta x_k$, where $k$ is the furthest site from $i$.
    Each inset shows $\w_{\uparrow}(\vect x)$  plotted along the black dashed line indicated in the main plot.
    Panel (c) shows the distribution of local energies for four interaction strengths. 
    Dashed lines represent power-law fits of the form $y=\frac{\alpha c}{(x_0-x)^{\alpha+1}}$. 
    The resulting characteristic exponents are shown in the figure and the tail coefficients are $0.7\pm0.4,4.1\pm1.5,15.4\pm3.5,(1.8\pm0.4)\times 10^2$ for $U=-2,-3,-4,-8$, respectively.
    The system is a $6\times 6\times 2$ honeycomb lattice Hubbard model at half filling. The site $i$ is $(5,5,A)$, while $j=(5,4,B)$ and  $k=(2,2,A)$ ($A$ and $B$ are sublattice indices).
    Each histogram in (c) is from $10^7$ data points.     
    Each $\w_{\uparrow}(\vect x)$ is normalized by its maximum value within the plotted region in (a) and (b).
    }
    \label{fig:honeycomb}
\end{figure}

Note that $w_{\uparrow}$ can always change sign: in sign-problem-free situations, $w_{\uparrow}$ and $w_{\downarrow}$ change sign at the same location, while in the presence of a sign problem, the locations are different which then leads to ``super-nodes'' in determinant space~\cite{shiweiZTCP1995,shiweiFTCP1999}.
As a result, the Boltzmann weight $w(\vect x)$ can vanish at certain configurations, 
which is the origin of the infinite variance problem.
These zeros form domain walls in auxiliary-field space, which can also lead to ergodicity problems in hybrid Monte Carlo or branching random walk algorithms \cite{white1988prb_ergodicity,shihao2013prb_symmetrycp,mingpu2016prb_ghfcp}. 
Consider the variance in the estimator in \Eq{O-est}:
\eq{
\text{Var}(O_{\text{loc}}) =  \frac{\sum_{\vect x} w(\vect x)O^2_{\text{loc}}(\vect x)}{\sum_{\vect x}w(\vect x)} - \bar O^2.
}{oloc_variance}
In the numerator, we can rewrite: $w(\vect x)\,O^2_{\text{loc}}(\vect x)=\frac{\bra{\phi(\vect x_L)}\hat O \ket{\phi(\vect x_R)}^2}{w(\vect x)}$ which clearly diverges as $w(\vect x)$ approaches zero. If this divergence is sufficiently strong, the variance becomes infinite. 

In \Fig{fig:honeycomb}, we examine the behavior of $w_\uparrow(\vect x)$ in the vicinity of a 
super-node or domain wall.
A nodal configuration is identified by detecting a sign flip in $w_\uparrow(\vect x)$ during the simulation.
The results reveal the nodal structure around this configuration. 
The insets further demonstrate that $w_\uparrow(\vect x)$ approaches zero linearly near the node.
This linear behavior has long been relied upon and validated in constrained path calculations~\cite{shiweiZTCP1995,shiweiFTCP1999}.
In the context of the infinite variance problem, the behavior of the weight near its zeros is directly linked to the characteristic exponent governing the distribution of observables.
If $w_\uparrow (\vect x)$ (or $w_\downarrow (\vect x)$) vanishes linearly with distance in the vicinity of the domain wall, the distribution of the local energy follows $\mathbb P(E_\text{loc}) \propto 1/|E_{\text{loc}}|^{5/2}$, 
as discussed in Ref.~\cite{assaad2022spikes}, 
yielding a power-law tail with characteristic exponent $\alpha = \frac 3 2$.
Our fitting results in the previous section are in good agreement with this prediction.
To further investigate the behavior of the characteristic exponent, we analyze a different lattice geometry—the honeycomb lattice—at various interaction strengths, as shown in panel (c) of \Fig{fig:honeycomb}. The resulting distributions again exhibit a power-law tail consistent with $\alpha = \frac 3 2$. 
At weaker interactions, we observe a reduction in the tail coefficient $c$, indicating fewer outlier events. Consequently, a larger number of samples is required to reliably extract the exponent in this regime. 
We conjecture that while the tail coefficient is sensitive to the lattice geometry and interaction strength, the characteristic exponent itself remains robust.
As a generalization, the SU($2N$) attractive Hubbard model will give a characteristic exponent $\alpha = 1+\frac 1 {2N}$ for calculating the $2N$-body correlation, which will be confirmed by numerical results (see Appendix~\ref{appendix:su4}).

In the discussion above, we have restricted to models in which the Boltzmann weight factorizes into several separated components.
For models with spin-orbit coupling or those belonging to the Majorana sign-problem-free class~\cite{Li2016PRL,han2024pfaffian}, the argument can be generalized.
It should also be emphasized that the characteristic exponent generally depends on the observable being measured.
In particular, when the local observable involves multiple Wick contraction terms, the divergent terms can cancel with each other or interfere in different ways, and care is needed in the analysis.

\subsection{The exact bridge link method} \label{subsec:exact_bridge_link}

Since the infinite variance problem arises from the zeros of the weight function, a natural idea is to eliminate these zeros \cite{shihao2016}, for example, by reducing the dimensionality of their manifold. 
The challenge is to do so while keeping changes to $w(\vect x)$ at a minimum, both for maintaining sampling efficiency (since in \Eq{O-est} it is near optimal for most quantities to sample $w(\vect x)$) and to meet the practical---yet extremely important---goal of limiting modifications to existing algorithms and codes.
The bridge link method \cite{shihao2016} is built on this insight.
Rather than sampling from the original weight $w(\vect x) = \langle\phi(\vect x_L)|\phi(\vect x_R)\rangle$, the auxiliary fields are sampled from a modified overlap that includes an inserted imaginary-time step, $\tilde w(\vect x) = \langle\phi(\vect x_L)|e^{-\delta \tau \hat H}|\phi(\vect x_R)\rangle$. 
Observables are then calculated using importance sampling:
\eq{\braket{O_{\text{loc}}}_{x\sim w(\vect x)} = \frac{\braket{\frac{w(\vect x)}{\tilde w(\vect x)}O_{\text{loc}}}_{x\sim \tilde w(\vect x)}}{\braket{\frac{w(\vect x)}{\tilde w (\vect x)}}_{x\sim \tilde w(\vect x)}}.}{}
We introduce the bridge operator $\hat \Lambda$, which denotes the operator inserted at the center of the imaginary-time propagation—specifically, $\hat \Lambda = e^{-\delta \tau \hat H }$ in the original bridge link method.
With this notation, the expectation value can equivalently be expressed as:
\eq{
\braket{O_{\text{loc}}}_{x\sim w(\vect x)} = \frac{\braket{\frac{O_{\text{loc}}}{\Lambda_{\text{loc}}}}_{x\sim \tilde w(\vect x)}}{\braket{\frac{1}{\Lambda_{\text{loc}}}}_{x\sim \tilde w(\vect x)}}.
}{bridge_exp}

In the bridge link method, sampling from the modified distribution $\tilde w \propto \langle\phi_L| e^{-\delta\tau \hat H}|\phi_R\rangle$ 
is performed by inserting another time step in the conventional DQMC simulation,
and then discarding the extra dimension, i.e., the additional
auxiliary-field associated with the time slice $^{-\delta\tau \hat H}$, in the evaluation in Eq.~(\ref{bridge_exp}).
It is easy to see that $\tilde w$ remains sign-problem-free and removes 
the infinite variance problem, as it can be written as 
\eq{
\tilde w(\vect x) = \sum_{x_{\text{bridge}}} \langle\phi(\vect x_L)|e^{-\frac{\delta\tau} 2 \hat K}e^{\hat v(x_{\text{bridge}})}e^{-\frac{\delta\tau} 2 \hat K}|\phi(\vect x_R)\rangle,
}{eq:bridge-value}
where we have applied the HS transformation, leading to the additional auxiliary-fields $x_{\text{bridge}}$. 
Since the model is sign-problem-free, every term in the sum is positive semi-definite.
As a result, $\tilde w(\vect x)$ can only vanish if all terms in the sum vanish simultaneously — a condition that imposes a much more restrictive constraint.
This dramatically reduces the dimensionality of the zero set, thereby suppressing the incidence of field configurations that contribute to infinite variance.

In Eq.~(\ref{bridge_exp}), we need to calculate 
$1/\Lambda_{\text{loc}}(\vect x)$ for each path, 
$\Lambda_{\text{loc}}=\frac{\langle\phi_L| e^{-\delta\tau \hat H}|\phi_R\rangle}{\left<\phi_L|\phi_R\right>}$
(omitting the auxiliary-field variables ${\vect x}$ for brevity).
This can be done by the Taylor expansion \eq{\Lambda_{\text{loc}}=1-\delta \tau \langle\hat H\rangle + \frac 1 2 \delta \tau^2 \langle\hat H^2\rangle+\cdots.}{} 
The computational cost is low if the expansion converges quickly to the desired accuracy, but 
the Taylor expansion will fail when $\braket{\hat H}
$ or higher-order terms become large, which is precisely the regime where the infinite variance problem typically occurs.
The previous method addresses this by introducing a secondary Monte Carlo to evaluate the integral/sum in Eq.~(\ref{eq:bridge-value}) when this situation arises.
We will refer to this approach as the sub-MC method. 
In this method, the expectation value ${\langle \Lambda_{\text{loc}}(\vect x})\rangle_{\text{sub-MC}}$ is computed for each DQMC path $\vect x$ for which 
$|E_\text {loc}(\vect x)-E_{\text{ref}}|$ exceeds a predefined threshold,
by  Monte Carlo integration over $x_{\text{bridge}}$. 
The sub-MC integration introduces additional errors or even a bias if the number of samples in the Monte Carlo integration is small. 
Although in practice this  can be mitigated by increasing the number of samples (for relatively low cost compared to the primary simulation) 
or by using the unbiased estimator~\cite{Alexandru2023prd}, these approaches are not perfect solutions due to the requirement of a secondary Monte Carlo procedure and independent sampling.

\begin{figure}
    \centering
    \includegraphics[width=1.0\linewidth]{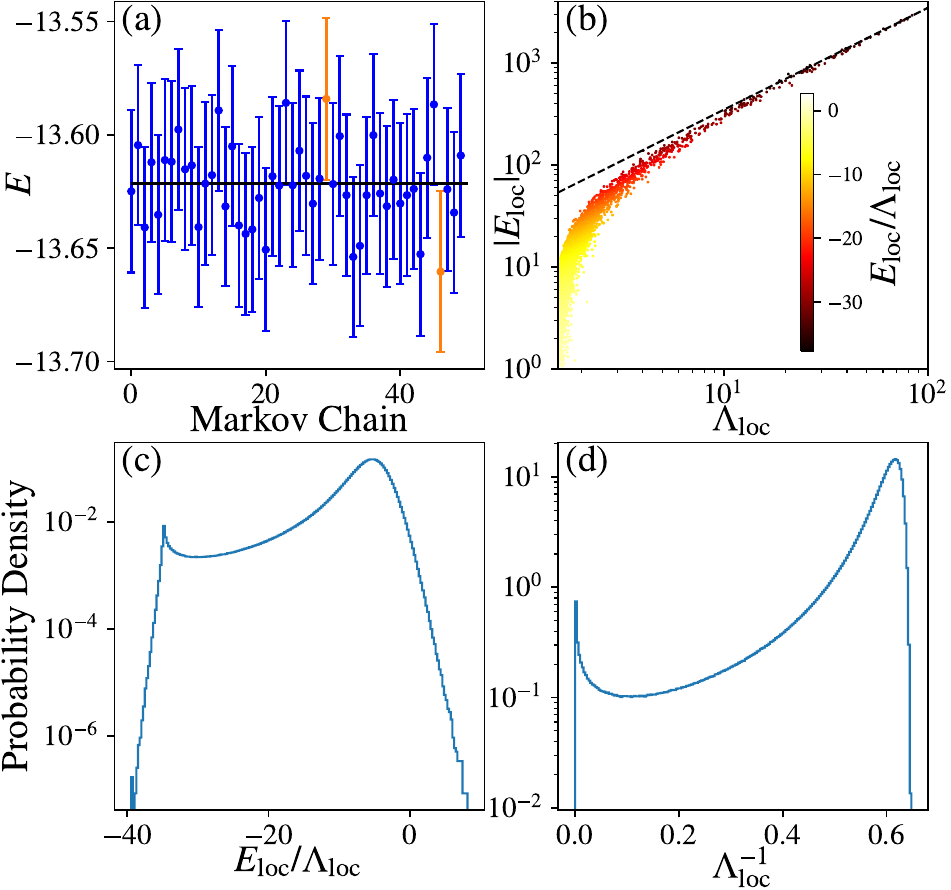}
    \caption{Illustrative results from the exact bridge link method.
    (a) DQMC results for the same system as in \Fig{fig:energy_without_bridge}(a).
    Note that the vertical axis scale has been shrunk by more than $15\times$ from \Fig{fig:energy_without_bridge}(a).
    Identical simulation parameters are used, except that the bridge link method is employed with bridge operator parameter $\theta = 1$ (\Eq{bridge_op}). 
    Error bars are estimated using the Jackknife method.
    (b) Scatter plot of $|E_{\text{loc}}|$ versus $\Lambda_{\text{loc}}$. While both quantities can diverge in magnitude, their ratio converges to a constant, indicated by the dashed line. 
    (c),(d) Probability densities of the ratio $E_{\text{loc}}/\Lambda_{\text{loc}}$ and of $\Lambda_{\text{loc}}^{-1}$, respectively.
    }
    \label{fig:bridge_data}
\end{figure}

The key advantage of the bridge link method lies in its minimal modification to the underlying quantum Monte Carlo sampling procedure. 
Therefore, it is desirable to develop an approach that eliminates the need for evaluating complicated integrals and any potential systematic error associated with it, while preserving the core idea of introducing a bridge link.
In the present work, we leverage the sign-problem-free nature of the model to construct bridge operators whose expectation values can be computed exactly in polynomial time, without requiring additional rounds of Monte Carlo sampling or introducing systematic error.
We refer to this approach as the {\textit{exact bridge link method}}.

To illustrate the idea of the new method, we begin with an example based on the attractive Hubbard model, which belongs to the Kramers sign-problem-free class \cite{congjun2005prb}.
We define an exact bridge operator \eq{
\hat \Lambda(\theta) = \frac 1 {N_{\text{site}}} \sum_{i} \left( \sum_{x = \pm 1} \exp(\mathrm i\theta x (\hat n_{i\uparrow}- \hat n_{i\downarrow}))\right),
}{}
where $\theta \in (0, \pi)$ is the bridge parameter.
The modified distribution $\tilde w(\vect x)$ using this bridge operator can also be efficiently sampled by introducing an additional time slice with a single auxiliary field that varies both in value and location $i$.
Importantly, there are only $2N_{\text{site}}$ terms in the sum, the bridge value $\Lambda_{\text{loc}}$ can be calculated exactly in $O(N_{\text{site}})$ time complexity, thereby resolving the issue present in the original bridge link method.

It is straightforward to verify that each term in the bridge operator respects the time-reversal symmetry $i\sigma_y K$ that governs the sign-free property of the attractive Hubbard model. 
Consequently, the expectation value of the operator is positive definite for any $\theta \in [0, \pi)$. 
This ensures that $\Lambda_{\text{loc}}(\theta) \geq 1 + \cos(\theta)$, and the variance of $1/\Lambda_{\text{loc}}$ is finite and bounded above by $1/(2 + 2\cos(\theta))$. 
Moreover, for the bridge operator itself, the expectation value defined in \Eq{bridge_exp} becomes $1/\braket{1/\Lambda_{\text{loc}}}_{x\sim \tilde w(\vect x)}$, which entirely avoids the infinite variance problem.
In fact, the bridge operator can be rewritten as
\eq{
\hat \Lambda(\theta) = 1 + \cos(\theta) + \frac{4-4\cos(\theta)}{N_{\text{site}}} \sum_i (\hat n_{i\uparrow}-\frac 1 2)(\hat n_{i\downarrow}-\frac 1 2),
}{bridge_op}
which is proportional to the interaction term in the Hubbard model. 
As a result, the interaction energy estimated using this bridge operator is also free from the infinite variance problem. 
The kinetic energy, involving only one-body operators, exhibits a much slower divergence; thus, the total energy also avoids infinite variance.

In \Fig{fig:bridge_data}, we apply this exact bridge operator with $\theta = 1$ to recalculate the same model presented in \Fig{fig:energy_without_bridge}.
As shown in \Fig{fig:bridge_data}(a), the true value lies within the error bars for nearly all independent runs, and no extreme deviations are observed.
In \Fig{fig:bridge_data}(b), we observe that both $E_{\text{loc}}$ and $\Lambda_{\text{loc}}$ diverge at the same rate, and their ratio saturates to the expected value $UN_{\text{site}} / (4 - 4\cos\theta)$, consistent with \Eq{bridge_op}.
The probability density plots of $E_{\text{loc}} / \Lambda_{\text{loc}}$ and $\Lambda_{\text{loc}}^{-1}$, shown in \Fig{fig:bridge_data}(c) and (d), respectively, confirm that these quantities are bounded. 
This demonstrates that the infinite variance problem is eliminated when using the exact bridge link method.

We now describe the exact bridge link method in a more general form. 
For models that are sign-problem-free due to the presence of two anti-unitary time-reversal symmetries $T_1,T_2$ \cite{Li2016PRL}, we can construct the exact bridge operator $\hat \Lambda$ as a sum of sign-problem-free operators. 
Specifically, the operator $\hat \Lambda$ is defined as:
\eq{
\hat \Lambda=\sum_{i=1}^{m} e^{\hat A}e^{\hat v_i} e^{\hat B}, 
}{def-bridge-T}
where $m$ is the number of the sign-free operators, and $\hat v_i$ are one-body operators with $O(1)$ support. 
Then the bridge value can be calculated in $O(m)$ time. 
In Eq.~(\ref{def-bridge-T}), the operators ${\hat A}$, ${\hat B}$, and ${\hat v_i}$ are all required to satisfy the same symmetries as the decomposed form of $e^{-\delta\tau \hat H}$, i.e., commute with $T_1$ and $T_2$, so as to maintain the sign-problem-free nature of the path. 
The main steps of the exact bridge link method are outlined as follows:
\begin{enumerate}
    \item Perform a standard DQMC simulation for all regular time slices.
    \item When reaching the bridge time slice, compute the propagated wavefunctions: $\ket{\phi_R'} = e^{\hat{B}}\ket{\phi_R}$,
    $\ket{\phi_L'} = e^{\hat{A}^\dagger}\ket{\phi_L}$ and evaluate the overlap ratio: $r = \frac{\braket{\phi_L'|\phi_R'}}{\braket{\phi_L|\phi_R}}$.
    \item Construct the Green's function from $\ket{\phi_{L,R}'}$ , and use the matrix determinant lemma to compute the local bridge weights: $w_\text{loc}(i)= \frac{\braket{\phi_L'|e^{\hat v_i}|\phi_R'}}{\braket{\phi_L'|\phi_R'}}$.
    \item Perform exact sampling of the index $i$ from the distribution $w_\text{loc}(i)$, and update the configuration accordingly.
    \item After completing a full sweep, return to the bridge slice and repeat Steps 2 and 3. The bridge value is then given by: $\Lambda_{\text{loc}}=r\times\sum_{i=1}^{m} w_{\text{loc}}(i)$. 
\end{enumerate}
The sign-free operators can be constructed to have strictly positive local expectation values, as in \Eq{bridge_op}, ensuring that $1/\Lambda_\text{loc}$ has finite variance. 
Consequently, the expectation values of such sign-free operators are inherently free from the infinite variance problem.
In practice, as long as $m$ is sufficiently large, the method remains applicable even when the sign-free operators can vanish occasionally.
The additional cost from the bridge link is proportional to $m$, and will remain well below the overall cost of the DQMC simulation unless $m=O(L_\tau N_{\text{site}}^3)$.
For example, in the earlier case where $\hat A = \hat B = 0$ and $ m = 2N_{\text{site}} $, the bridge link only adds a negligible fraction to the total computational complexity.

\begin{figure}
    \centering
    \includegraphics[width=1.0\linewidth]{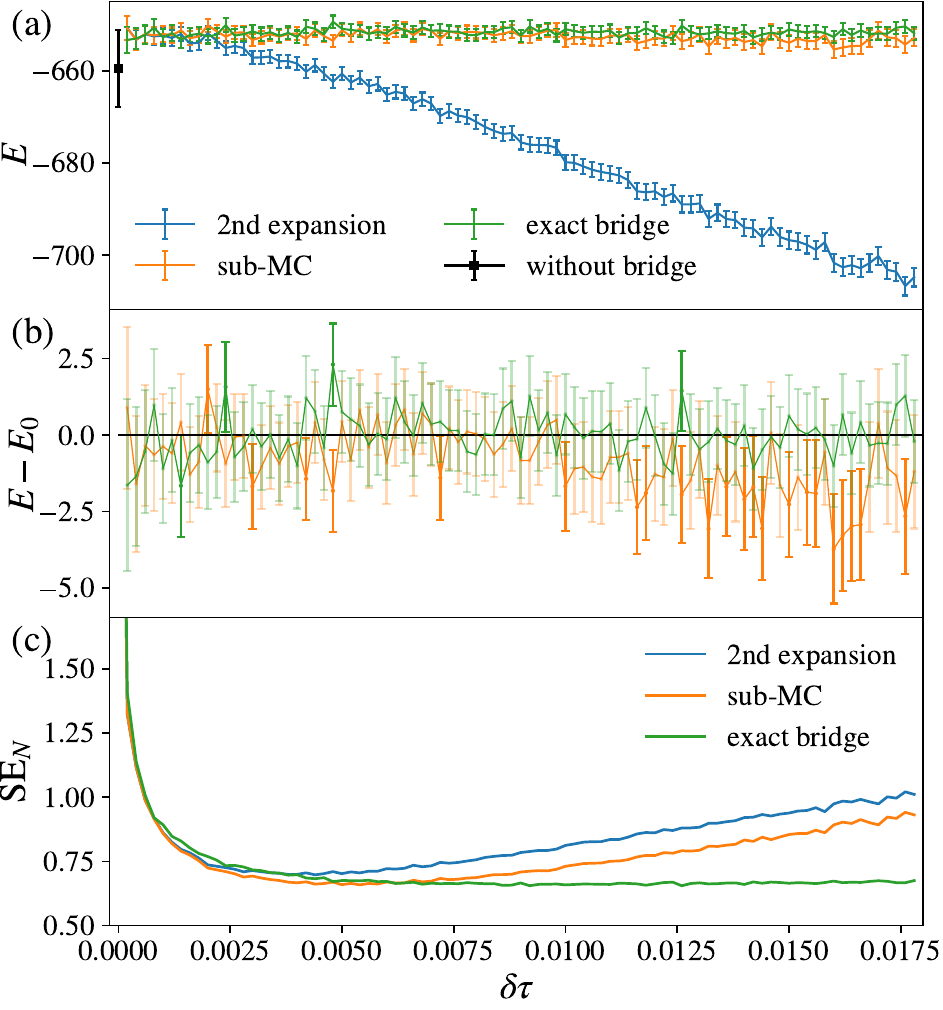}
    \caption{Comparison between the exact and original bridge link methods.
    (a) Energy obtained versus the size of the bridge link, $\delta\tau$. 
    (b) Zoom-in on the energy differences between sub-MC results, the exact bridge link result, and the overall sample mean ($E_0$) obtained from the exact bridge method. Data points where $E_0$ lies outside the error bar are shown with reduced transparency for emphasis.
    (c) Finite-sample standard error of the mean, $\text{SE}_N$, for different bridge methods and time steps.
    The system is the attractive Hubbard model on a $12 \times 12$ lattice with 144 electrons and interaction strength $U = -8$. 
    The simulation parameters are $\Theta=20,\Delta\tau=0.05$.
    Each data point represents $10^5$ Monte Carlo sweeps, with the first quarter discarded to ensure thermalization. 
    In the sub-MC calculations for the original bridge link method, $1000$ samples are used.
    }
    \label{fig:compare}
\end{figure}

Next, we compare the exact bridge link method with the original approach by introducing a modified bridge operator designed to approximate the time-evolution operator used in the original method. 
The modified bridge operator is defined as:
\eq{
\hat \Lambda'(\delta\tau) \propto \frac 1 {N_{\text{site}}} \sum_{i} \left( \sum_{x = \pm 1} e^{-\frac{\delta\tau }2\hat K}e^{\mathrm{i}\theta x (\hat n_{i\uparrow}- \hat n_{i\downarrow})}e^{-\frac{\delta\tau }2\hat K}\right),
}{}
where $\theta$ is chosen as
$\theta = \arccos\left( \frac{4 + N_{\text{site}} U \delta\tau}{4 - N_{\text{site}} U \delta\tau} \right)$, 
to approximate the evolution operator $e^{-\delta\tau \hat{H}}$ used in the original bridge link method.
In \Fig{fig:compare}, we compare the exact and original bridge link methods. 
The problem studied is the attractive Hubbard model on a larger $12 \times 12$ lattice with a stronger interaction strength of $U = -8$, providing a stringent test case with a pronounced 
infinite variance problem.
By varying the bridge time step $\delta\tau$, we observe that the original method—based on a second-order expansion—exhibits significant deviation as $\delta\tau$ increases.
To improve accuracy, we also employ the sub-MC approach with 1000 samples to estimate the bridge value in the original method. 
While this greatly improves the results, a close-up view as in \Fig{fig:compare}(b) still reveals a systematic bias that grows with increasing $\delta\tau$.
In \Fig{fig:compare}(c), we plot the SEM as a function of $\delta\tau$. For the original method, the SEM initially decreases with increasing $\delta\tau$ but then increases again.
In contrast, the exact bridge method exhibits much more stable behavior across a wide range of time steps.

It is useful to note the behavior of the SEM at small $\delta\tau$.
Even with a tiny bridge step of $\delta\tau = 0.0002$, the SEM drops dramatically to 1.4.  
(The nominal estimate of SEM for $\delta\tau = 0$, i.e., no bridge, is 4.2, which is of course incorrect and should be enlarged by applying the rescaling factors discussed in Sec.~\ref{subsec:error_estimation}.)
This is due to the fundamental difference in scaling behavior between methods with and without a bridge operator — namely, the presence of infinite variance in the latter. 
As sample size increases, this improvement becomes even more pronounced.

\subsection{Application to the SU(4) Hubbard model: charge-$4e$ correlations} \label{subsec:SU4}

We next present a real application of the exact bridge link method to a challenging problem, namely 
the attractive SU(4) Hubbard model. 
This model is of particular interest for studying quartet condensation, also known as charge-$4e$ superconductivity \cite{berg2009charge4e,shaokai2021charge4e,ashvin2022symmetry,wangjian2024prx,Herland2010,samoilenka2025microscopic}.  Unlike the conventional charge-$2e$ superconducting state, characterized by a pair order parameter involving two creation operators, 
the charge-$4e$ state is described by the quartet order parameter.
Computing the charge-$4e$ correlation function is challenging due to both infinite variance and a large noise-to-signal ratio — particularly because the long-range correlation signal may remain small even in the ordered phase. As we illustrate below, the new method makes it possible to perform ``clean'' simulations to tackle this problem.

\begin{figure}
    \centering
    \includegraphics[width=0.9\linewidth]{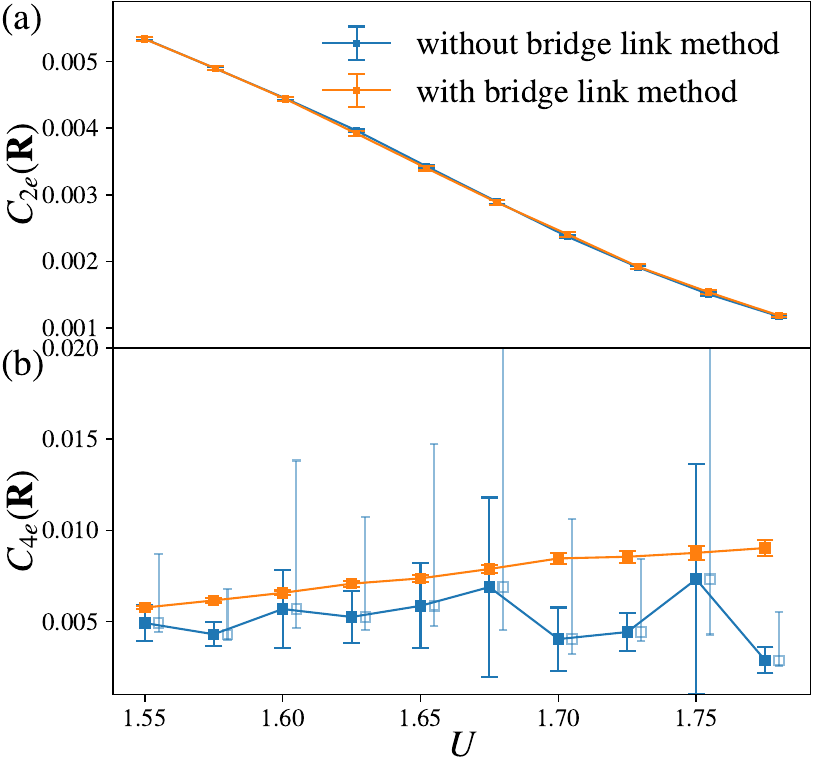}
    \caption{Charge-$2e$ and charge-$4e$ correlations in the attractive SU(4) model. 
    $C_{2e}({\mathbf R})$ (panel a) and $C_{4e}({\mathbf R})$ (panel b) correlations are computed and shown for the largest distance $\mathbf{R} = (L/2,L/2)$, for a range of interaction strengths $U$, with and without the bridge link method.  
    For charge-$4e$ correlation results obtained without the bridge link, two types of error bars are shown: one based on $2 \times \text{SE}_N$ and a lighter-colored bar reflecting a rescaled estimate incorporating the characteristic exponent $\alpha = 1.25$ (see Appendix~\ref{appendix:su4}).
    The system is a $10 \times 10$ lattice with 25 electrons per flavor. 
    Simulations use $\Theta = 100$, $\Delta\tau = 0.1$, and the bridge link operator in \Eq{SUNbridge} with parameter $\lambda = 5$. Each point averages 10 independent runs of $3\times10^5$ sweeps, with the first half discarded.
    }
    \label{fig:su4}
\end{figure}

The SU(4) Hubbard Hamiltonian is given by
\eq{\hat H= -t \sum_{\alpha,\langle ij\rangle}{c_i^\alpha}^\dagger c_j^\alpha +\text{h.c.}  -\frac U 2 \sum_{i}(\sum_{\alpha}\hat n_i^\alpha)^2,
}{}
where $\alpha = 1, 2, 3, 4$ is the flavor index. In our simulations, we employ the exact HS transformation introduced in Ref.~\cite{wu2014prl}.
The charge-$4e$ state is described by the quartet order parameter $\Delta_{4e}(i)\equiv c_i^{1\dagger} c_i^{2\dagger} c_i^{3\dagger} c_i^{4\dagger}$, and we compute the corresponding correlation function $C_{4e}({\mathbf r})=\sum_i [\Delta_{4e}^\dagger(i)\,\Delta_{4e}(j)+\text{h.c.}]/N_\text{site}$, with sites $j$ and $i$ connected by the lattice vector ${\mathbf r}$. 
Similarly, the charge-$2e$ correlation is defined using the pair order parameter $\Delta_{2e}(i)\equiv c_i^{1\dagger} c_i^{2\dagger}$, without loss of generality.

The bridge link operator we use here is
\eq{
\hat \Lambda''(\lambda) = \frac 1 {N_{\text{site}}} \sum_{i}
\left( e^{\lambda \sum_\alpha c^{\alpha\dagger}_i c^\alpha_j}
+e^{\lambda \sum_\alpha c^{\alpha\dagger}_j c^\alpha_i}\right),
}{SUNbridge}
where $i,j$ are sites connected by vector $\mathbf{R} = (L/2, L/2)$, the maximum spatial separation on an $L \times L$ lattice.
The corresponding bridge expectation value is given by 
\eq{
\Lambda''_{\text{loc}}(\lambda)=\frac 1 {N_{\text{site}}}\sum_i\left((1-\lambda G_{ij})^4
+(1-\lambda G_{ji})^4\right),
}{}
which regularizes any divergence arising from the Green’s functions $G_{ij}$ and $G_{ji}$. 
Here, we omit the flavor index since all flavors are identical in our simulation.
Note that the local expectation value of charge-$4e$ correlation is calculated as $G_{ij}^4+G_{ji}^4$, so using $\hat \Lambda''(\lambda)$ as the bridge operator guarantees that this observable has finite variance.

As shown in \Fig{fig:su4}, simulations without the bridge link method result in large fluctuations in the charge-$4e$ correlation. In contrast, the charge-$2e$ correlation exhibits much smaller fluctuations. Indeed, as discussed in Appendix~\ref{appendix:su4}, they have different characteristic exponents and the charge-$2e$ correlation in this case does not have an infinite variance problem. 
The exact bridge link method preserves the efficiency for charge-$2e$ correlation while eliminating the infinite variance problem for charge-$4e$ correlation. 
The latter is now determined with significantly reduced statistical error.
We observe clearly that the charge-$4e$ order parameter increases over the range of interaction strengths studied, while the charge-$2e$ order parameter decreases. 
This contrasting behavior suggests a possible quantum phase transition between the charge-$2e$ and charge-$4e$ superconducting phases. 
A detailed investigation of this phase transition will be presented elsewhere.

\section{Discussion and Concluding Remarks} \label{sec:discussion}
In this work, we have presented a comprehensive analysis of the infinite variance problem in DQMC simulations. 
We identified that the infinite variance problem arises from the emergence of heavy-tailed distributions in certain observables, leading to  statistical uncertainties  scaling with sample size as 
$\epsilon \propto N^{\frac{1}{\alpha}-1}$.
This is markedly slower than the standard $1/\sqrt{N}$ behavior, and 
causes conventional error estimation to systematically underestimate the confidence interval.
We proposed a retrospective error remediation strategy based on a rescaled SEM. 
The rescaling factors depend solely on the properties of the tail distribution — specifically, the characteristic exponent $\alpha$ and the skewness parameter $\beta$.
We observed that the characteristic exponent $\alpha$ can vary across different systems and observables.
In Hubbard-like models, numerical results and theoretical analysis show excellent consistency.
However, in other systems, such as those belonging to the Majorana sign-problem-free class, the tail behavior may differ qualitatively.
Developing a broader and unified understanding of how the exponent $\alpha$ depends on model structure, symmetry class, and observable type would be an important direction for future research. 
Even without knowledge of the tail parameters, we presented a strategy to obtain tail-aware error estimation with sufficient sample size, which is straightforward to adopt in DQMC as a standard post-analysis procedure.

To eliminate the infinite variance problem completely, we proposed the exact bridge link method, which represents a significant advancement over the original bridge link approach. 
In particular, the bridge operator itself can be shown to have finite variance. 
As long as the bridge value diverges at the same rate or faster than the observable of interest, the infinite variance problem can, in principle, be resolved.
In contrast to the original bridge link method, in which a sub-MC estimator is needed that may introduce larger computational cost or even systematic bias, the exact bridge link method avoids such approximation altogether. 
Importantly, the method is straightforward to incorporate into existing DQMC frameworks, requiring only minimal modifications and introducing negligible computational overhead. 
This makes it a practical and robust solution for enhancing the reliability of DQMC simulations.

There are many areas in which the exact bridge operator can potentially lead to progress, and merit further exploration.
Leveraging sign-problem-free operators belonging to lower symmetry classes \cite{wei2018semigroup} may offer advantages in specific systems.
The bridge operator may also provide a promising strategy for overcoming ergodicity barriers in HMC simulations.
While our current formulation is restricted to sign-problem-free models, generalizing the exact bridge link method to systems with a sign problem is possible, and warrants further investigation \cite{ryan2024}. 
Straightforward applications of the method to computations of time-displaced observables can result in a substantial increase in computational complexity and will require additional development.

As we discussed in the introduction, the infinite variance problem is present in many other QMC or machine learning/VMC algorithms. 
We believe that the analysis of heavy-tailed distributions presented in this work, along with the proposed general idea of the exact bridge link method for importance sampling, may find broader applications in these contexts.

\section{Acknowledgments}
The Flatiron Institute is a division of the Simons Foundation. The authors would like to
thank Shao-Kai Jian, Ze-Yao Han, Xuan Zou, Yiqi Yang, Chunhan Feng, and Ryan Levy for helpful discussions.

\newtheorem{definition}{Definition}
\appendix
\renewcommand{\thefigure}{S\arabic{figure}}
\setcounter{figure}{0}

\section{Stable distribution and GCLT.}
\label{appendix}
This section provides background on the generalized central limit theorem and stable distributions, both of which are essential for understanding the heavy-tailed distribution discussed in the main text.
Key results are summarized below; for a comprehensive treatment, we refer readers to Refs.~\cite{gnedenko1968limit,nolan2020univariate}.

We begin by defining stable distributions, which generalize the normal distribution and naturally emerge as the limiting distributions of normalized sums of independent and identically distributed (i.i.d.) random variables with heavy tails.
\begin{definition}
A random variable $X$ is said to follow a stable distribution if its characteristic function takes the form
\eq{
\mathbb{E} \exp(i uX) = 
\exp{\left(i\delta u -|\gamma u|^\alpha(1-i\beta\text{sgn}(u) \Phi(u))\right)},
}{}
where $\Phi(u) = \begin{cases}
    \tan(\frac{\pi\alpha}{2})&\alpha \neq 1\\
    -\frac 2 \pi \log|u|&\alpha = 1
\end{cases},
$ $0<\alpha\leq 2,-1\leq \beta \leq 1, \gamma> 0, \delta \in \mathbb R$. The distribution is denoted as $\mathbf S(X;\alpha,\beta,\gamma,\delta)$ and is referred to as a \textbf{stable distribution}.
\end{definition}

The stable distribution is also referred to as Lévy alpha-stable distribution, named after Paul Lévy, who first studied this class of distributions nearly a century ago \cite{lévy1925calcul}.
Here, we adopt Nolan’s first parameterization of stable distributions \cite{nolan2020univariate}. A stable distribution is characterized by four parameters: the characteristic exponent $\alpha$;
the skewness parameter $\beta$;
the scale parameter $\gamma$;
and the location parameter $\delta$.
When $\alpha\neq 1$, the standardized form satisfies $(X-\delta)/\gamma\sim \mathbf{S}(\alpha,\beta,1,0)$.
The mean of the distribution exists and is equal to $\delta$ when $\alpha>1$.
Stable distributions are so named because the sum of i.i.d. random variables drawn from a stable distribution also follows a stable distribution with the same $\alpha$.
For $\alpha<2$, the stable distribution is heavy-tailed with infinite variance, whereas for $\alpha = 2$, the stable distribution reduces to the normal distribution.

\begin{definition}
A sequence of random variables $\{X_n\}$ is said to converge in distribution to a random variable $X$, denoted as:
\eq{X_n\stackrel{d}\rightarrow X,}{}
if for all points $x$ at which the cumulative distribution function (CDF) $F_X(x)=\mathbb P(X\leq x)$ is continuous,
\eq{\lim_{n\rightarrow\infty} F_{X_n}(x)=F_X(x),}{}
where $F_{X_n}(x)$ is the CDF of $X_n$.
\end{definition}

The GCLT characterizes the conditions under which a normalized sum of i.i.d. variables converges in distribution to a stable distribution.

\begin{theorem}[Generalized Central Limit Theorem]
Let $X_1, X_2,\dots$ be i.i.d. copies of $X$ where $X$ has characteristic function $\Phi_X(u)$ and satisfies the tail conditions:
\eq{
\begin{array}{l}
   \lim_{x\rightarrow\infty}x^\alpha F_X(-x) =  c^- , \\
   \lim_{x\rightarrow\infty} x^\alpha (1-F_X(x)) =  c^+ ,
\end{array}
}{}
where $c^-\geq 0, c^+\geq 0$ and $0<c^-+c^+<\infty$.
Then the normalized sum \eq{a_n(X_1+X_2+\cdots + X_n)-b_n\stackrel{d}{\rightarrow}Z,}{} with $Z\sim\mathbf{S}(\mathrm{min}(2,\alpha),\beta,1,0)$ and
\begin{widetext}
\eq{
\begin{array}{llll}
  \beta = \frac{c^+-c^-}{c^++c^-}  &, a_n = \left(\frac{2\Gamma(\alpha)\sin(\frac{\pi\alpha}{2})}{\pi(c^++c^-)}\right)^{1/\alpha} n^{-1/\alpha}&, b_n = 0 & \text{if \ \  } 0<\alpha <1,\\
  \beta = \frac{c^+-c^-}{c^++c^-}&, a_n = \left(\frac{2\Gamma(\alpha)\sin(\frac{\pi\alpha}{2})}{\pi(c^++c^-)}\right)^{1/\alpha} n^{-1/\alpha}&, b_n = n\mathrm{Im} \log\Phi_X(a_n) & \text{if \ \  } \alpha = 1,\\
    \beta = \frac{c^+-c^-}{c^++c^-}  &, a_n = \left(\frac{2\Gamma(\alpha)\sin(\frac{\pi\alpha}{2})}{\pi(c^++c^-)}\right)^{1/\alpha} n^{-1/\alpha}&, b_n = na_n\mathbb E[X]&\text{if \ \  } 1<\alpha < 2,\\
    \beta = 0 &, a_n = \left((c^++c^-)n\log n\right)^{-1/2}&, b_n = na_n\mathbb E[X]&\text{if \ \  } \alpha = 2,\\
    \beta = 0 &, a_n = \left(n \sigma^2_X/2\right)^{-1/2}&, b_n = na_n\mathbb E[X]&\text{if \ \  } \alpha > 2.\\
\end{array}
}{}
\end{widetext}
\end{theorem}
\textit{Proof.} For a detailed proof, see Ref.~\cite[pg.~175]{gnedenko1968limit} and Ref.~\cite[pg.~142]{nolan2020univariate}.

\begin{corollary}
Let \( F_X(x) \) be a heavy-tailed distribution with characteristic exponent \( \alpha \in (1, 2) \). Then, as \( N \to \infty \), the following scaling relations hold in distribution:
\begin{equation}
\begin{array}{llll}
&\frac{(\bar X_N -\mathbb E [X])}{N^{1 /\alpha-1}}&\stackrel{d}{\rightarrow} \eta_\alpha(c^-+c^+)Z &\text{with }Z\sim\mathbf{S}(\alpha,\beta,1,0), \\
&\frac{s^2_N(X) }{N^{2/\alpha-1}}&\stackrel{d}{\rightarrow} \eta_{\frac \alpha 2}(c^-+c^+)  Z' &\text{with }Z'\sim\mathbf{S}(\frac \alpha 2,1,1,0), \\

\end{array}
\end{equation}
where $\eta_\alpha(c) = \left(\frac{2\Gamma(\alpha)\sin(\frac{\pi\alpha}{2})}{\pi c}\right)^{-1/\alpha}$.
\end{corollary}

\textit{Proof.} The first result follows directly from Theorem 1 in the case where $1<\alpha<2$.  To derive the second result, consider the distribution of $Y = X^2$.  
The cumulative distribution function of $Y$ is given by:
\eq{
F_Y(y) = 
\begin{cases}
F_X(\sqrt{y}) - F_X(-\sqrt{y}), & y > 0, \\
0, & y \leq 0.
\end{cases}
}{}
Analyzing the tail behavior of  $F_Y(y)$, it is easy to find as $y\rightarrow \infty$,
$y^{\alpha/2}(1-F_Y(-y)) \rightarrow 0$ and $y^{\alpha/2}(1-F_Y(y)) = \sqrt{y}^\alpha(1-F_X(\sqrt y )+F_X(-\sqrt y)) \rightarrow c^++c^-$, 
which implies that $c^-{}'=0, c^+{}' = c^-+c^+$.
From Theorem 1, we know that:
\eq{{N^{1-\frac 2 \alpha}}\bar Y_N \stackrel{d}{\rightarrow}  \eta_{\frac \alpha 2}(c^-+c^+)Z',}{}
with $Z'\sim\mathbf{S}(\alpha/2,1,1,0)$.

Next, we consider the sample variance $s^2_N(X) \equiv \frac{N}{N-1}(\bar Y_N-\bar X_N^2)$, in the limit of $N\rightarrow \infty$, $\bar X_N^2$ becomes negligible after scaling by $1/N^{2/\alpha-1}$.
Thus, the sample variance converges to the desired limiting form after scaling by $1/N^{2/\alpha-1}$.
As a direct consequence, the SEM also converges after scaling as:
\eq{\frac{\text{SE}_N(X)}{N^{1/\alpha -1}} \stackrel{d}{\rightarrow} \sqrt{\eta_{\frac \alpha 2}(c^-+c^+)Z'}, }{}
where $Z'\sim\mathbf{S}(\alpha/2,1,1,0)$.
Although the standard error also converges to a heavy-tailed distribution, the characteristic exponent is $\alpha$, which is the same as that of $F_X(x)$,  ensuring that the distribution has a well-defined mean.
From the scaling form of $\frac{\text{SE}_N(X)}{N^{1/\alpha -1}}$, we can observe that the standard error estimator correctly captures the scaling of the fluctuations of the sample mean.

\section{Correlation distribution in attractive SU(4) Hubbard model} \label{appendix:su4}

\begin{figure}
    \centering
    \includegraphics[width=1.0\linewidth]{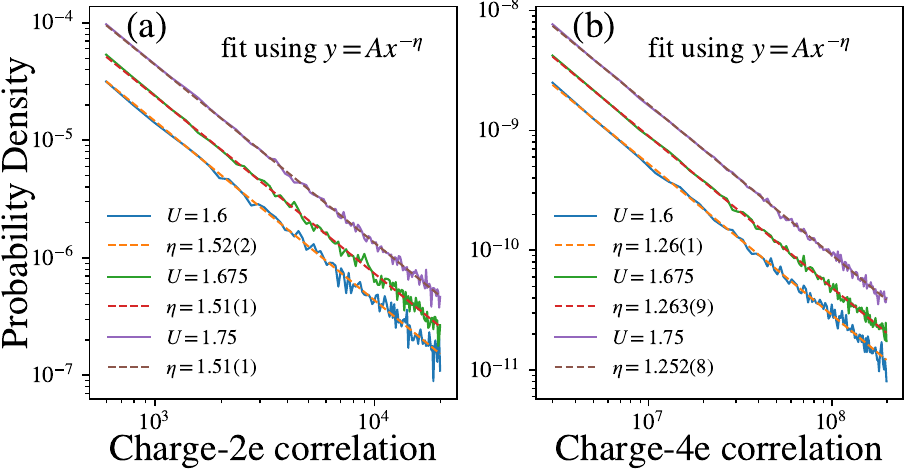}
    \caption{(a) Distribution of the local expectation value of the charge-$2e$ correlation (with bridge link), fitted using the function $y = Ax^{-\eta}$.
    (b) Distribution of the local expectation value of the charge-$4e$ correlation (with bridge link), also fitted using $y = Ax^{-\eta}$.
    }
    \label{fig:su4_2}
\end{figure}

As a direct generalization of the discussion in Section~\ref{subsec:dqmc_and_ivp}, we can extend the prediction of characteristic exponents to the attractive SU($2N$) Hubbard model. 
In this case, when approaching the nodal line, the determinant of a single flavor is expected to vanish linearly with the perpendicular distance $x_\bot$ from a configuration $\vect{x}$ to the nodal line: $w^\alpha(\vect x)\propto x_{\bot}$,
where $\alpha = 1, \dots, 2N$ denotes the flavor index. 
In this regime, the Green's function elements diverge as $G_{ij} \propto 1/x_\bot$. 
The total weight scales as $w(\vect x)=[w^\alpha(\vect x)]^{2N}\propto x_\bot^{2N}$.
If an observable is computed as $O_\text{loc} = G_{ij}^m \propto x_\bot^{-m}$ for some integer $m$—for instance, $m = 2$ for a pair correlation and $m = 4$ for a quartet correlation—then the probability distribution of the local observable behaves as $\mathbb P(O_\text{loc})\propto |O_\text{loc}|^{-1-\frac{2N+1}{m}}$, leading to a characteristic exponent $\alpha = \frac{2N+1}{m}$.
Specifically, in the SU(4) case, the charge-$4e$ correlation is associated with a characteristic exponent $\alpha = 5/4$, indicating a more severe infinite variance problem than in the SU(2) case. In contrast, the charge-$2e$ correlation has a larger exponent $\alpha = 5/2 > 2$, which places it outside the infinite variance regime.

As shown in \Fig{fig:su4_2}(a) and (b), the tail distribution fits reveal that the charge-$4e$ correlation exhibits a fitted exponent $\eta_{4e}\simeq1.25$ with the bridge link, while the charge-$2e$ correlation yields $\eta_{2e} \simeq 1.5$. 
Taking into account the role of the importance function, this corresponds to characteristic exponents of $\alpha = \eta_{4e} = 1.25$ for charge-$4e$ and $\alpha = \eta_{2e} + 1 = 2.5$ for charge-$2e$ correlations in the absence of a bridge operator, in good agreement with theoretical predictions.
The characteristic exponent appears relatively insensitive to interaction strength but can vary significantly depending on the specific model and observable under study.

\bibliography{ref}

\end{document}